\documentclass[
 reprint,
 amsmath,amssymb,
 aps,
]{revtex4-2}

\usepackage{graphicx}
\usepackage{subcaption}
\usepackage{dcolumn}
\usepackage{bm}
\usepackage{physics}
\usepackage{esdiff}

\DeclareMathOperator{\diag}{diag}

\begin{document}
\preprint{APS/123-QED}

\title{Hybrid Quantum-Classical Normalizing Flow}
\author{Anlei Zhang$^1$}
\author{Wei Cui$^2$}
\email{aucuiwei@scut.edu.cn}

\address{$^1$School of Physics and Optoelectronics, South China University of Technology, Guangzhou 510640, China}
\address{$^2$School of Automation Science and Engineering, South China University of Technology, Guangzhou 510640, China}

\begin{abstract}
With the rapid development of quantum computing technology, we have entered the era of noisy intermediate-scale quantum (NISQ) computers. Therefore, designing quantum algorithms that adapt to the hardware conditions of current NISQ devices and can preliminarily solve some practical problems has become the focus of researchers. In this paper, we focus on quantum generative models in the field of quantum machine learning, and propose a hybrid quantum-classical normalizing flow (HQCNF) model based on parameterized quantum circuits. Based on the ideas of classical normalizing flow models and the characteristics of parameterized quantum circuits, we cleverly design the form of the ansatz and the hybrid method of quantum and classical computing, and derive the form of the loss function in the case that quantum computing is involved. We test our model on the image generation problem. Experimental results show that our model is capable of generating images of good quality. Compared with other quantum generative models, such as quantum generative adversarial networks (QGAN), our model achieves lower (better) Fréchet inception distance (FID) score, and compared with classical generative models, we can complete the image generation task with significantly fewer parameters. These results prove the advantage of our proposed model.
\end{abstract}

\maketitle

\section{\label{Introduction}Introduction}
With the continuous development and progress of society, people's demand for computing power is growing. However, Moore's Law, which describes the development of classical electronic computing, is difficult to maintain due to the size of transistors gradually approaching the physical limit. This has prompted people to do research on new computing technologies. Among them, quantum computing has received much attention in recent years due to its enormous potential.

The hardware technology for quantum computing is booming, with some milestone achievements such as Sycamore \cite{arute2019quantum} quantum computer by Google and Jiuzhang \cite{zhong2020quantum} quantum computer by the University of Science and Technology of China. However, the existing quantum computing hardware is still in the early stages of development, with a small number of qubits and high error rate, making it difficult to support the quantum algorithms designed by researchers in the past that require a large number of qubits and low error rates to run well on perfect quantum hardware. At the same time, although some existing experiments can prove that quantum computers have quantum advantages over classical computers in specific tasks, there is still a long way to go before surpassing classical computers in practical problems. Therefore, based on the hardware conditions of the currently available Noisy Intermediate-Scale Quantum (NISQ) \cite{preskill2018quantum} computers, developing algorithms the can run on current quantum hardware and solve some problems with practical significance has become the focus of research.

Machine learning, especially deep learning \cite{lecun2015deep}, plays an increasingly important role in the development of artificial intelligence. Due to the enormous resources required for classical computers to perform neural network computations, designing machine learning algorithms that run on quantum computers and utilizing their unique properties is receiving increasing attention \cite{schuld2015introduction, biamonte2017quantum, massoli2022leap, wittek2014quantum, schuld2018supervised, rebentrost2014quantum, schuld2021machine}. Considering the respective characteristics and advantages of quantum computers and classical computers, many research efforts consider designing hybrid quantum-classical algorithms to take advantage of current quantum hardware. Quantum circuits are composed of quantum gates, which are different from the gate circuits in classical computing. The gate in quantum computing may contain parameters, enabling their effects to vary continuously. The proposal of the variational quantum eigensolver algorithm for molecular ground states has inspired the design of variational quantum algorithms \cite{cerezo2021variational, fedorov2022vqe} and the design of algorithms based on parameterized quantum circuits \cite{benedetti2019parameterized}.

In the field of classical deep learning, researchers have proposed many excellent models for generative tasks, such as Variational Autoencoders (VAE) \cite{kingma2014auto}, Generative Adversarial Networks (GAN) \cite{goodfellow2014generative, heusel2017gans}, Normalizing Flows \cite{dinh2015nice, dinh2017density}, Diffusion Models \cite{ho2020denoising}, etc., which have achieved good results in areas like image generation. The GAN model consists of a generator and a discriminator. The expected function of the generator is to sample from noise and transform the noise to obtain generated data. The expected function of the discriminator is to accept a data sample and determine whether the data is real data from the training set or data generated by the generator. During the training process, the training objective of the generator is to maximize the probability that the generated images are judged as real data by the discriminator. The training objective of the discriminator is to distinguish as much as possible which images are generated by the generator and which images are real data from the training set. Normalizing Flows are a type of reversible neural network design. During the training process, it learns how to transform the images in the training set into a standard distribution (such as a standard normal distribution) through reversible transformations. During generation, it samples from the selected standard distribution and performs the learned reversible transformation in reverse to generate images. However, although these classical generative models can achieve good results, the cost of training and generation is high, and further improvement of their performance is limited by the computational power of classical computing hardware. Moreover, classical computing devices face many difficulties in further improving computational power due to the transistor size approaching physical limits. Therefore, in this paper, we design quantum generative models to explore the use of quantum computing to accomplish generative learning tasks.

Currently, there exist many research works on quantum generative models \cite{tian2023recent}, mainly including Quantum Circuit Born Machine (QCBM) \cite{liu2018differentiable}, Quantum Generative Adversarial Network (QGAN) \cite{lloyd2018quantum, dallaire2018quantum, zoufal2019quantum, huang2021experimental, hu2019quantum}, Quantum Boltzmann Machine (QBM) \cite{amin2018quantum}, and Quantum AutoEncoder (QAE) \cite{romero2017quantum, khoshaman2018quantum}. In the QGAN model, researchers take the basic principles of classical generative adversarial networks as a starting point. They use parameterized quantum circuits as the implementation of the generator, while the discriminator and parameter update process are implemented using classical computers. However, the existing quantum generative models generally face difficulties in fitting nonlinear transformations and have insufficient expressive power, which hinders the application of quantum generative models to solve more complex real-world problems. 

To address this problem, we design and implement a hybrid quantum-classical normalizing flow model with stronger expressive power that can learn more complex mappings. We draw on the ideas of classical normalizing flows and cleverly design a new way quantum and classical computing are combined. We design our model based on parameterized quantum circuits and rigorously derive the loss function which undergoes changes in quantum setting. We choose PyTorch and PennyLane frameworks to implement the our proposed model and train it on MNIST \cite{lecun2010mnist} dataset to examine the image generation ability of the model. In terms of experimental results, qualitative results show that our proposed model can generate data that is similar to the training data, and quantitative results show that our model's FID score is better than the QGAN model. We further analyze the qubits and parameters requirement of our model and make comparisons with classical generative models. These results demonstrate the effectiveness of our model design and prove the advantages of the HQCNF model we propose for the first time.

This paper is organized as follows. Section \ref{Preliminaries} are preliminaries, in which we will review the classical normalizing flow model and introduce the data encoding strategy. Our proposed hybrid quantum-classical model will be presented in Section \ref{Normalizing Flow in Quantum Computation}. In Section \ref{Numerical Results}, we present numerical simulation that supports our proposals. Conclusion and discussion are presented in Section \ref{Conclusion}.

\section{\label{Preliminaries}Preliminaries}
\subsection{Classical Normalizing Flow}

Normalizing flows are a powerful generative model in the field of classical machine learning, first proposed by Dinh et al. \cite{dinh2015nice, dinh2017density}. We use multidimensional vectors to represent samples (for example, all pixel values of an image are expanded to form a multidimensional vector). Generally speaking, the task of a generative model is to find the probability distribution that the real data follows. To express this probability distribution, the model starts with a random vector $Z$ that follows a simple probability distribution as input, and through a transformation $f_\theta(\cdot)$ determined by the model parameters $\theta$, obtains a vector $X=f_\theta(Z)$ that follows a complex probability distribution as output. The task of model training is to make the probability distribution followed by the output vector $X$ as close to the probability distribution followed by the real data as possible. Thus, by sampling samples $z$ from the simple input probability distribution and obtaining $x=f_\theta(z)$ through the model transformation, we obtain data generated by the model that is difficult to distinguish from real data.

Let the probability density function of the probability distribution followed by the random vector $Z$ be $\pi(z)$, and the probability density function of the probability distribution followed by the generated data $X$ be $p_\theta(x)$. According to the transformation formula of probability density function when variables are changed, 
	\begin{eqnarray}
	p_\theta(x)=\frac{\pi(z)}{|\det \bm J|}, 
	\end{eqnarray}
where
	\begin{eqnarray}
	\bm{J} =
		\begin{pmatrix}
		\frac{\partial x_1}{\partial z_1} & \cdots & \frac{\partial x_n}{\partial z_1} \\
		\vdots & \ddots & \vdots \\
		\frac{\partial x_1}{\partial z_n} & \cdots & \frac{\partial x_n}{\partial z_n}
		\end{pmatrix}
	\end{eqnarray}
is the Jacobian matrix of $f_\theta$, supposing both $X$ and $Z$ are $n$ dimensional vectors.

The probability density function of the probability distribution followed by the output random vector $X$ is the parametric $p_\theta(x)$. Seeking the parameter $\theta$ that makes the probability distribution followed by $X$ as close to the probability distribution followed by the real data as possible is a parameter estimation problem. We use the maximum likelihood estimation method to estimate the parameter $\theta$. Let the training dataset be $D=\{x^{(1)}, x^{(2)}, …, x^{(m)}\}$, then the likelihood function is
	\begin{eqnarray}
	L(\theta) 
	= \prod_{i = 1}^m p_\theta(x^{(i)}) 
	= \prod_{i = 1}^m \frac{\pi(z)}{|\det \bm J_\theta^{(i)}|} 
	= \prod_{i = 1}^m \frac{\pi(f_\theta^{-1}(x^{(i)}))}{|\det \bm J_\theta^{(i)}|}.
	\end{eqnarray}
Our goal is to find $\arg \max_\theta L(\theta)$.

We use gradient-based methods, which is widely adapted in deep learning, to find the optimal parameter $\theta$. We define the loss function as 
	\begin{eqnarray} \label{eq:loss}
	\begin{split}
	\mathcal{L}(\theta)
	&= -\log L(\theta) \\
	&= -\sum_{i=1}^m \log \pi(f_\theta^{-1}(x^{(i)}) + \sum_{i=1}^m \log |\det \bm J_\theta^{(i)}|
	\end{split}
	\end{eqnarray}
so our goal is to find $\arg \min_\theta \mathcal{L}(\theta)$. From the expression of $\mathcal{L}(\theta)$, it can be seen that in order to make the expression efficiently computable, the transformation function of the model is required to be invertible, and its Jacobian determinant should be easy to calculate. These two points are the key to designing normalizing flow models.

Dinh et al. proposed a clever method for designing bijective functions $f_\theta(\cdot)$ in their paper \cite{dinh2017density}. The structure of a layer in this model is as follows. For simplicity, the following description takes a single sample as an example, and the principle is the same for batch input of multiple samples. Let the layer index be $l$, the input data be $x^{[l-1]}$, and the output be $x^{[l]}$, both with dimension $D$. This method first divides the input vector $x^{[l-1]}$ into two parts: $x_{1:d}^{[l-1]}$ and $x_{d+1:D}^{[l-1]}$, and then constructs the output vector $x^{[l]}$ of the function as
	\begin{eqnarray}
	x^{[l]}_{1: d} = x^{[l-1]}_{1: d}
	\end{eqnarray}
	\begin{eqnarray}
	x^{[l]}_{d+1: D}=x^{[l-1]}_{d+1: D} \bigodot e^{s_\theta^{[l]}(x^{[l-1]}_{1: d})} + t_\theta^{[l]}(x^{[l-1]}_{1: d}), 
	\end{eqnarray}
in which $s_\theta^{[l]}$ and $t_\theta^{[l]}$ are functions of the $l$-th layer, accepting the $1$st to the $d$-th (inclusive) variables of the $(l-1)$-th layer as input, and outputting a $(D-d)$-dimensional vector. $e^{s_\theta^{[l]}(x^{[l-1]}_{1: d})}$ represents taking the element-wise exponential of the $(D-d)$-dimensional vector output by $s_\theta^{[l]}$, and $\bigodot$ represents element-wise multiplication. The specific mapping relationships of $s_\theta^{[l]}$ and $t_\theta^{[l]}$ are determined by the learnable parameters within them (the parameters $\theta$ of the entire model are composed of the parameters of the $s$ function and $t$ function of each layer). They can have arbitrary functional forms, such as deep neural networks, and do not need to have invertible properties.

It is easy to notice that the mapping constructed above is invertible, namely
	\begin{eqnarray}
	x^{[l-1]}_{1: d} = x^{[l]}_{1: d}
	\end{eqnarray}
	\begin{eqnarray}
	x^{[l-1]}_{d+1: D} = (x^{[l]}_{d+1: D} - t_\theta^{[l]}(x^{[l]}_{1: d})) \bigodot e^{-s_\theta^{[l]}(x^{[l]}_{1: d})}.
	\end{eqnarray}
In addition, its Jacobian determinant is also easy to compute, with the complexity $O(D)$ instead of naively $O(D^3)$, since
	\begin{eqnarray}
	\bm J_\theta^{[l]} =
		\begin{pmatrix}
			\bm{I}_d & \cdots \\
			\bm 0_{d \times (D - d)} & \diag(e^{s_\theta^{[l]}(x^{[l-1]}_{1: d})})
		\end{pmatrix}
	\end{eqnarray}
leads to 
	\begin{eqnarray}
	\det \bm J_\theta^{[l]} = \prod_{j=1}^{D-d} e^{[s_\theta^{[l]}(x^{[l-1]}_{1: d})]_j}
	\end{eqnarray}
and
	\begin{eqnarray}
	\log |\det \bm J_\theta| = \sum_{l=1}^L \log |\det \bm J_\theta^{[l]}| = \sum_{l=1}^L \sum_{j=1}^{D-d} [s_\theta^{[l]}(x^{[l-1]}_{1: d})]_j.
	\end{eqnarray}
Subsitituting into Eq.~\eqref{eq:loss}, we can obtain the loss function.

The classical normalizing flow has achieved excellent results in various fields such as image generation \cite{dinh2015nice, dinh2017density} and quantum many-body system simulation \cite{dugan2023q, wang2023learning}. It is recognized as one of the best generative models. This paper will take the classical normalizing flow as a starting point and design a hybrid quantum-classical normalizing flow model based on parameterized quantum circuits.

\subsection{Quantum Data Encoding}
There are several ways to encode classical data into quantum circuits, such as basis encoding \cite{schuld2021machine}, amplitude encoding \cite{plesch2011quantum}, angle encoding and Hamiltonian encoding \cite{berry2007efficient}. Amplitude encoding and angle encoding are used in this paper.

To use amplitude encoding to encode a feature vector $\bm x$ of length $N$
	\begin{eqnarray}
	\bm x = 
	\begin{pmatrix}
	x_0 & x_1 & \cdots & x_{N - 1}
	\end{pmatrix}
	\end{eqnarray}
(assume $N$ a integer power of 2 for simplicity) into a quantum circuit with $n$ qubits ($N=2^n$), we normalize the feature vector to make sure equation
	\begin{eqnarray}
	\sum_{i = 0}^{N - 1} |x_i|^2 = 1
	\end{eqnarray}
is satisfied. Then, according to the procedure to prepare arbitrary quantum state, we prepare the state 
	\begin{eqnarray}
	\ket \psi = \sum_{i = 0}^{N - 1}{x_i \ket i}
	\end{eqnarray}
as the initial state of the quantum circuit. Thus, the classical features are encoded into quantum circuits.

To expliot angular encoding, we first notice that the evolution of quantum states follows the Schrodinger equation
	\begin{eqnarray}
	i\hbar \diff{\ket \psi}t = H \ket \psi
	\end{eqnarray}
thus
	\begin{eqnarray}
	\ket{\psi(t)} = e^{- i t H / \hbar} \ket{\psi(0)}.
	\end{eqnarray}
When the Hamiltonian is set fixed (for example, we may choose 
	\begin{eqnarray}
	H / \hbar = \sigma_y
	\end{eqnarray}
where
	\begin{eqnarray}
	\sigma_y = 
		\begin{pmatrix}
		0 & -i \\
		i & 0
		\end{pmatrix}
	\end{eqnarray}
), the time $t$ can be used as a parameter to encode classical data into quantum circuit. This method is used to encode the data of classical hidden layer into quantum circuit in this paper.

\section{\label{Normalizing Flow in Quantum Computation}Normalizing Flow in Quantum Computation}
In this section, we will introduce our method to implement normalizing flow in a hybrid quantum-classical way and the describe the structure of our proposed model. Inspired by the idea of classical normalizing flows, the hybrid quantum-classical normalizing flow makes use of the intrinsic invertibility of quantum computation that originates from the unitary evolution postulate of quantum mechanics. Like its classical counterpart, hybrid quantum-classical normalizing flow also consists of several layers. 
\begin{figure}[h!]
	\centering
	\begin{subfigure}[b]{1.0\linewidth}
		\includegraphics[width=\linewidth]{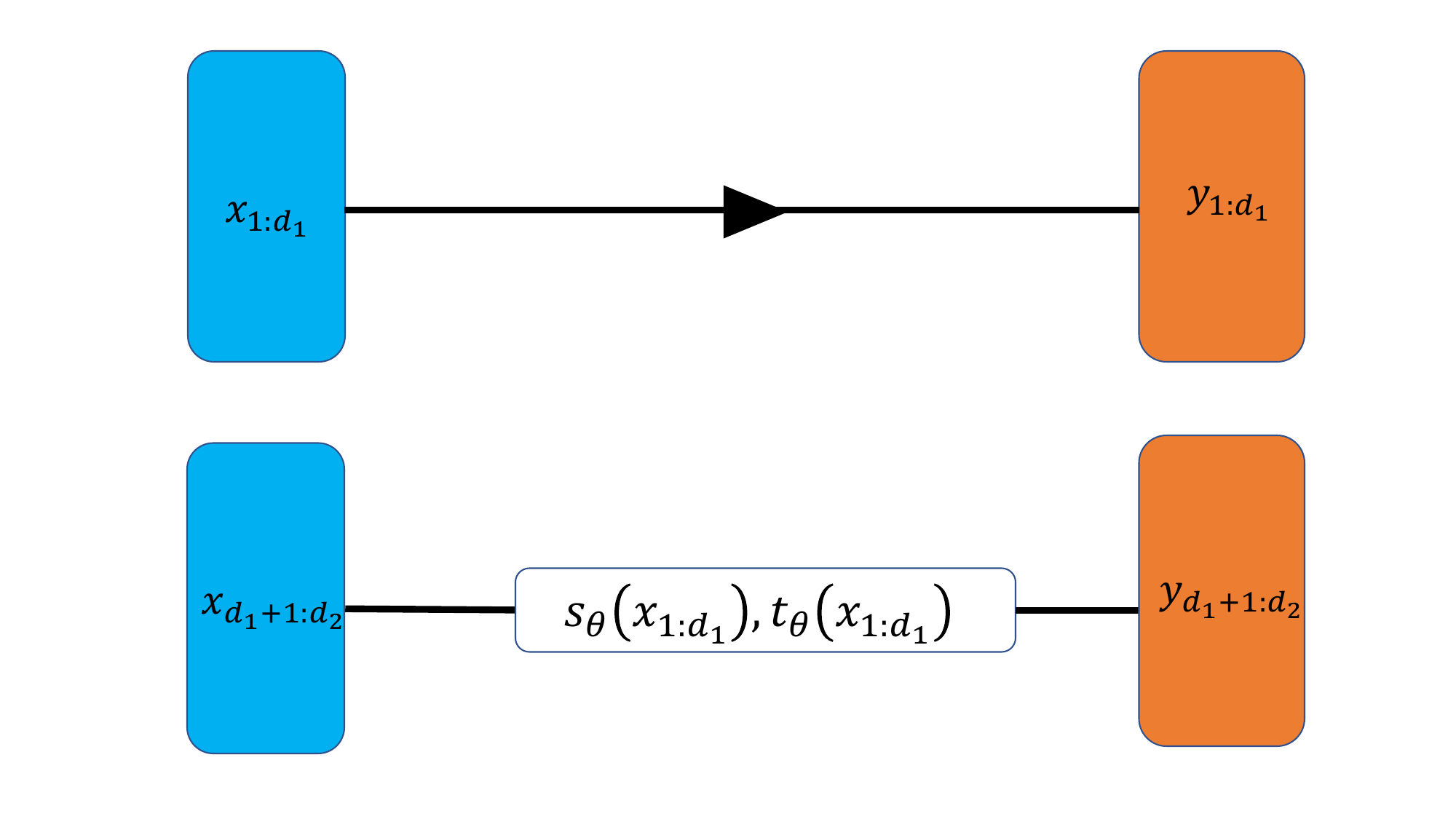}
			\caption{Classical part of the model.}
	\end{subfigure}
	\begin{subfigure}[b]{1.0\linewidth}
		\includegraphics[width=\linewidth]{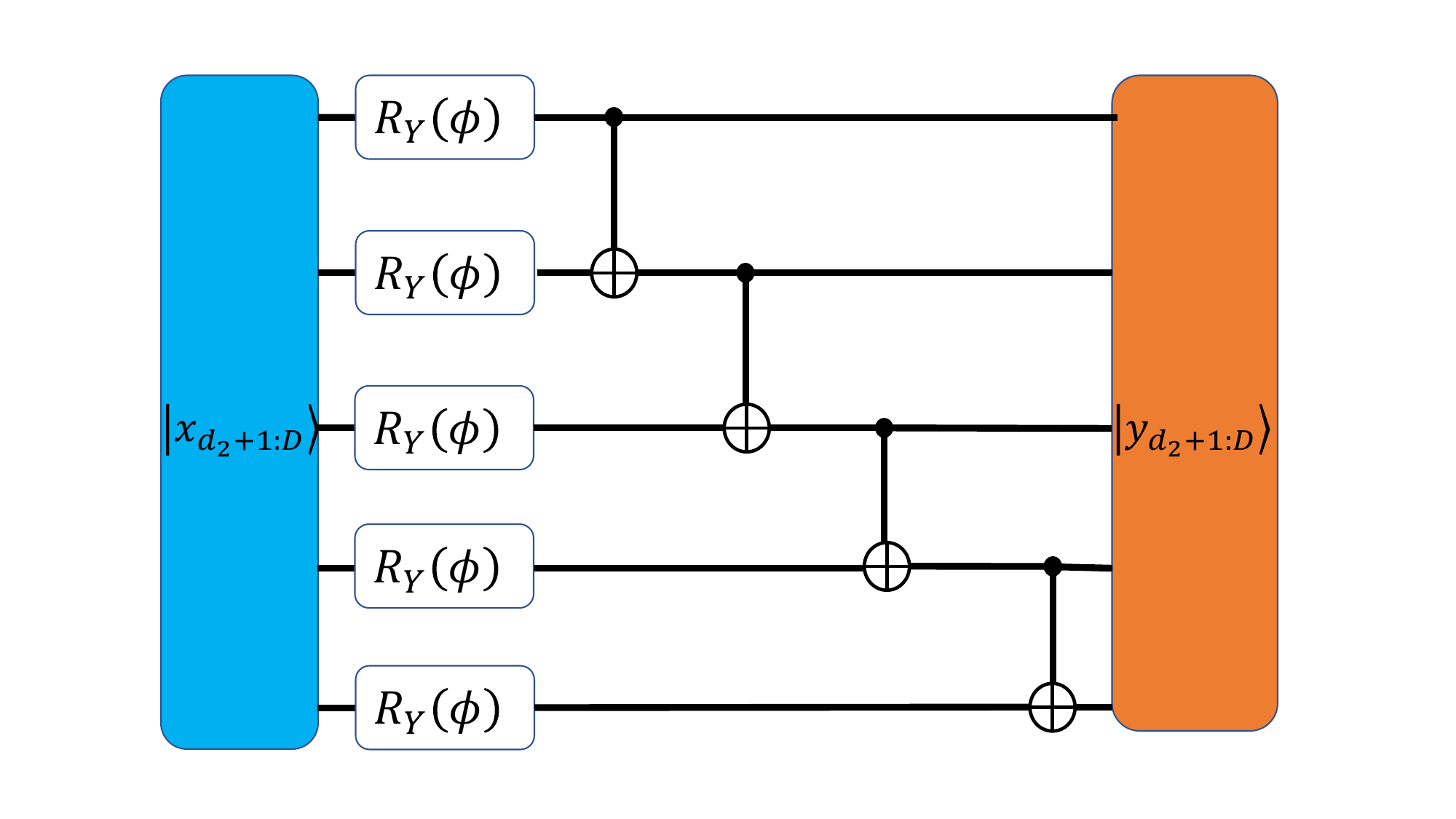}
			\caption{Quantum circuit of the model.}
	\end{subfigure}
	\caption{The model structure of hybrid quantum-classical normalizing flow.}
	\label{fig:model structure}
\end{figure}

The structure of one layer is shown below. The layer transforms vector $\bm x$ to vector $\bm y$ by performing a nonlinear bijection. The length of both $\bm x$ and $\bm y$ is $D$. We get $y_{1: d_1}$ by
	\begin{eqnarray}
	y_{1: d_1} = x_{1: d_1},
	\end{eqnarray}
and we get $y_{d_1 + 1: d_2}$ by
	\begin{eqnarray}
	y_{d_1 + 1: d_2} = x_{d_1 + 1: d_2} \bigodot \exp[s_\theta(x_{1: d_1})] + t_\theta(x_{1: d_1}),
	\end{eqnarray}
where $s_\theta$ and $t_\theta$ can be arbitrary (classical) neural networks, and $\bigodot$ means element-wise product. Another classical neural network $r_\theta$ is used to map $x_{1: d_1}$ to parameters $\phi$ in quantum circuit ansatz, namely
	\begin{eqnarray}
	\phi = r_\theta(x_{1: d_1}).
	\end{eqnarray}
And we get $y_{d_2 + 1: D}$ by 
	\begin{eqnarray}
	y_{d_2 + 1: D} = \bm U(\phi) x_{d_2 + 1: D},
	\end{eqnarray}
where $y_{d_2 + 1: D}$ and $x_{d_2 + 1: D}$ are amplitudes of the final and initail state of the quantum circuit and $\bm U(\phi)$ represents the evolution of the quantum state through the parameterized quantum circuit.

This map from $\bm x$ to $\bm y$ is invertible. We can get $\bm x$ back from $\bm y$ by
	\begin{eqnarray}
	x_{1: d_1} = y_{1: d_1}
	\end{eqnarray}
	\begin{eqnarray}
	x_{d_1 + 1: d_2} = (y_{d_1 + 1: d_2} - t_\theta(x_{1: d_1}))\bigodot \exp[-s_\theta(x_{1: d_1})]
	\end{eqnarray}
and
	\begin{eqnarray}
	x_{1: d_1} = [\bm U(\phi)]^{-1}y_{1: d_1 + 1}.
	\end{eqnarray}
Since $x_{1: d_1} = y_{1: d_1}$, $\phi$ can be got from $\bm y$ instead of $\bm x$, namely
	\begin{eqnarray}
	\phi = r_\theta(y_{1: d_1})
	\end{eqnarray}
for now. Moreover, $[\bm U(\phi)]^{-1}$ could be implemented efficiently, owing to the invertible nature of quantum circuit and quantum mechanics.

Next, we will work out the Jacobin determinant for this layer. Since it is easier to deal with real numbers than complex numbers, we treat $\Re[x_{d_2 + 1: D}]$ (the real part of $x_{d_2 + 1: D}$), $\Im[x_{d_2 + 1: D}]$ (the imaginary part of $x_{d_2 + 1: D}$), $\Re[y_{d_2 + 1: D}]$ (the real part of $y_{d_2 + 1: D}$) and $\Im[y_{d_2 + 1: D}]$ (the imaginary part of $x_{d_2 + 1: D}$) as independent variables, and we write $U$ as 
	\begin{eqnarray}
	U = V + iW,
	\end{eqnarray}
where $V$ and $W$ are both real. Thus, we have
	\begin{eqnarray}
	\begin{split}
	&\quad \Re[y_{d_2 + 1: D}] + i\Im[y_{d_2 + 1: D}] \\
	&= (V + iW)(\Re[x_{d_2 + 1: D}] + i\Im[y_{d_2 + 1: D}]) \\
	&= (V\Re[x_{d_2 + 1: D}] - W\Im[y_{d_2 + 1: D}]) \\
	&+ i(W\Re[x_{d_2 + 1: D}] + V\Im[y_{d_2 + 1: D}]).
	\end{split}
	\end{eqnarray}
It can be written in another way,
	\begin{eqnarray}
	\begin{pmatrix}
		\Re[y_{d_2 + 1: D}] \\
		\Im[y_{d_2 + 1: D}]
	\end{pmatrix}
	=
	\begin{pmatrix}
		V & -W \\
		W & V
	\end{pmatrix}
	\begin{pmatrix}
		\Re[x_{d_2 + 1: D}] \\
		\Im[x_{d_2 + 1: D}]
	\end{pmatrix},
	\end{eqnarray}
so the $[D + (D - d_2)] \times [D + (D - d_2)]$ dimensional Jacobian matrix is
	\begin{eqnarray}
	J=
	\begin{bmatrix}
	I & \cdots & \cdots \\
	0 & \diag \left[ e^{s_\theta(x_{1: d_1})} \right] & 0 \\
	0 & 0 &
		\begin{pmatrix}
				V & -W \\
				W & V
		\end{pmatrix}^T
	\end{bmatrix}.
	\end{eqnarray}
And according to Laplace expansion theorem, 
	\begin{eqnarray}
	\det J = 
		\det(\diag \left[ e^{s_\theta(x_{1: d_1})} \right])
		\det
			\begin{pmatrix}
				V & -W \\
				W & V
			\end{pmatrix}.
	\end{eqnarray}
So the key point now is to compute
	\begin{eqnarray}
	\det
		\begin{pmatrix}
			V & -W \\
			W & V
		\end{pmatrix}.
	\end{eqnarray}
Since $U$ is unitary, we have
	\begin{eqnarray}
	UU^\dagger = I
	\end{eqnarray}
namely
	\begin{eqnarray}
	\begin{split}
	&\quad (V + iW)(V^T - iW^T) \\
	&= VV^T + WW^T + i(WV^T - VW^T) \\
	&= I,
	\end{split}
	\end{eqnarray}
thus
	\begin{eqnarray}
	VV^T + WW^T = I
	\end{eqnarray}
and
	\begin{eqnarray}
	WV^T - VW^T = 0.
	\end{eqnarray}
So
\begin{eqnarray}
	\begin{pmatrix}
		V^T & W^T \\
		-W^T & V^T
	\end{pmatrix}
	\begin{pmatrix}
		V & -W \\
		W & V
	\end{pmatrix}
=
	\begin{pmatrix}
		I & 0 \\
		0 & I
	\end{pmatrix}
\end{eqnarray}
and finally we get
	\begin{eqnarray}
	\left[
		\det
			\begin{pmatrix}
			V & -W \\
			W & V
			\end{pmatrix}
	\right]^2
	=
	\det
		\begin{pmatrix}
			I & 0 \\
			0 & I
		\end{pmatrix}
	= 1,
	\end{eqnarray}
namely
	\begin{eqnarray}
	\det
		\begin{pmatrix}
			V & -W \\
			W & V
		\end{pmatrix}
	= 1.
	\end{eqnarray}
So
	\begin{eqnarray}
	\det J = \prod_j (\exp[s_\theta(x_{1: d_1})])_j.
	\end{eqnarray}

In our implementation, the hybrid quantum-classical normalizing flow are composed of several layers whose structures are shown in Figure \ref{fig:model structure}. The unchanged features in one layer ($y_{1: d_1} = x_{1: d_1}$) will be changed in the next layer. 

After training, we want our model to be able to transform a noise vector which is centered around ${\ket 0}^{\otimes n}$ to a handwritten digit, so we set $\pi_0(\cdot)$ in the loss function Eq.~\eqref{eq:loss} to estimate how close $f_\theta^{-1}(x)$ is to ${\ket 0}^{\otimes n}$; the closer it is, the smaller $\pi_0(f_\theta^{-1}(x))$ will be.

\section{\label{Numerical Results}Numerical Results}
\begin{figure}[h!]
	\centering
	\begin{subfigure}[b]{0.17\linewidth}
		\includegraphics[width=\linewidth]{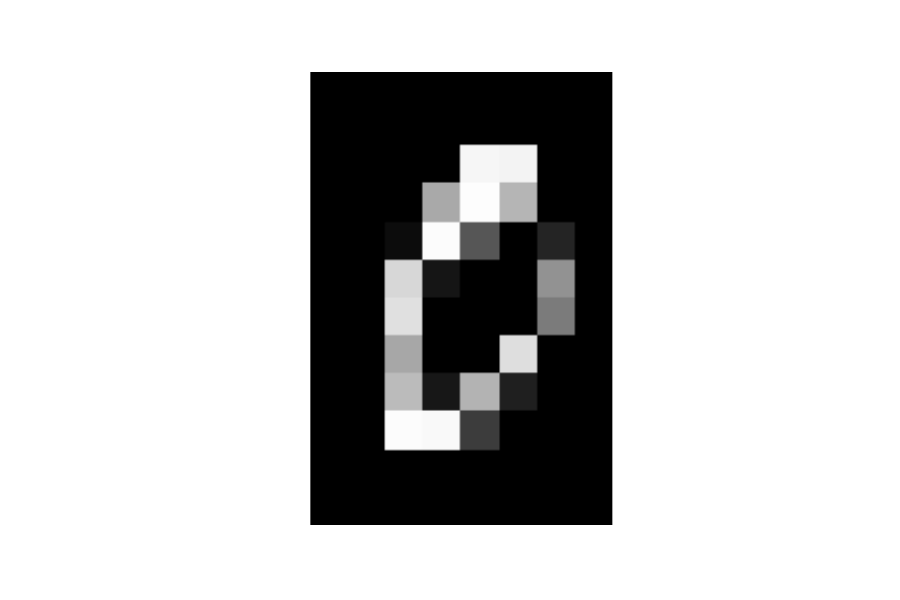}
			\caption{zero}
	\end{subfigure}
	\begin{subfigure}[b]{0.17\linewidth}
		\includegraphics[width=\linewidth]{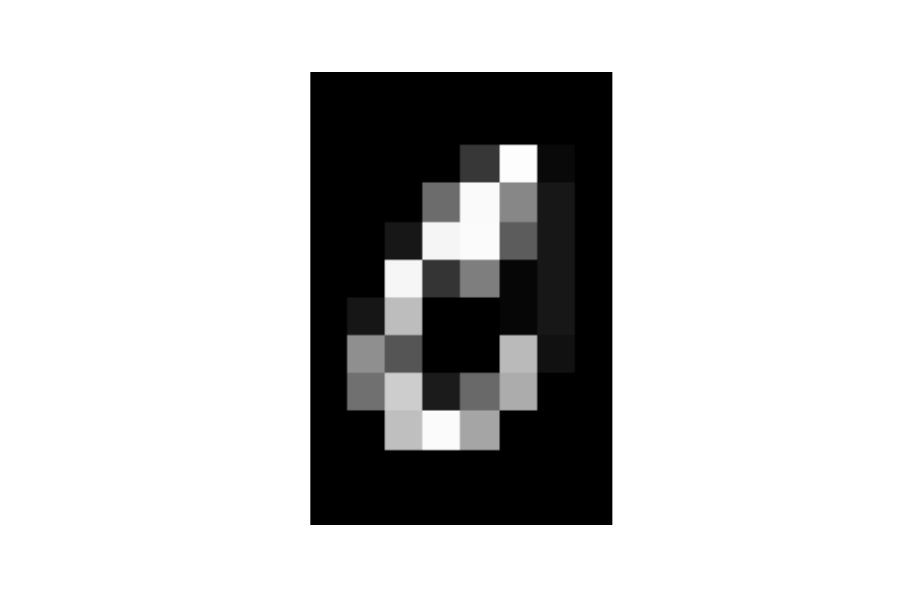}
			\caption{zero}
	\end{subfigure}
	\begin{subfigure}[b]{0.17\linewidth}
		\includegraphics[width=\linewidth]{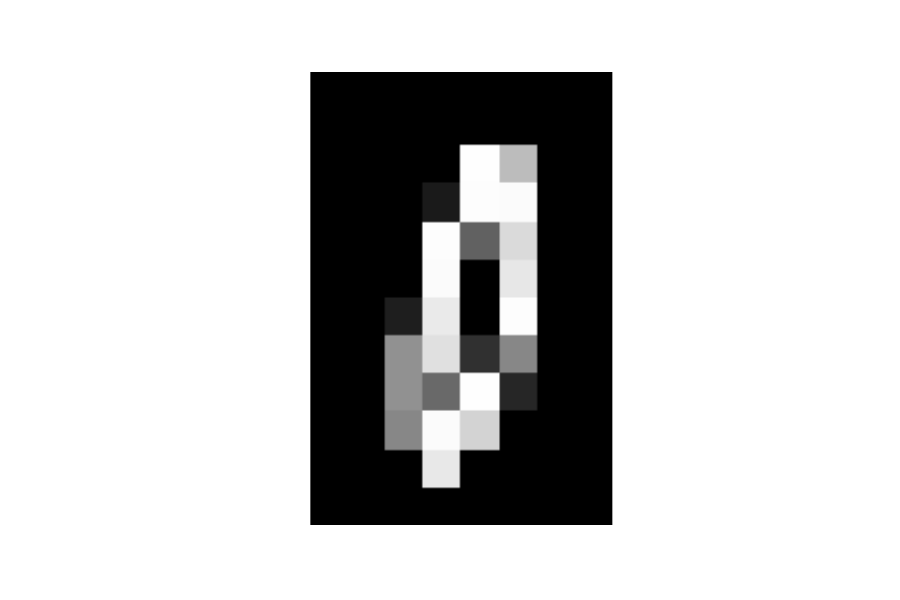}
			\caption{zero}
	\end{subfigure}
	\begin{subfigure}[b]{0.17\linewidth}
		\includegraphics[width=\linewidth]{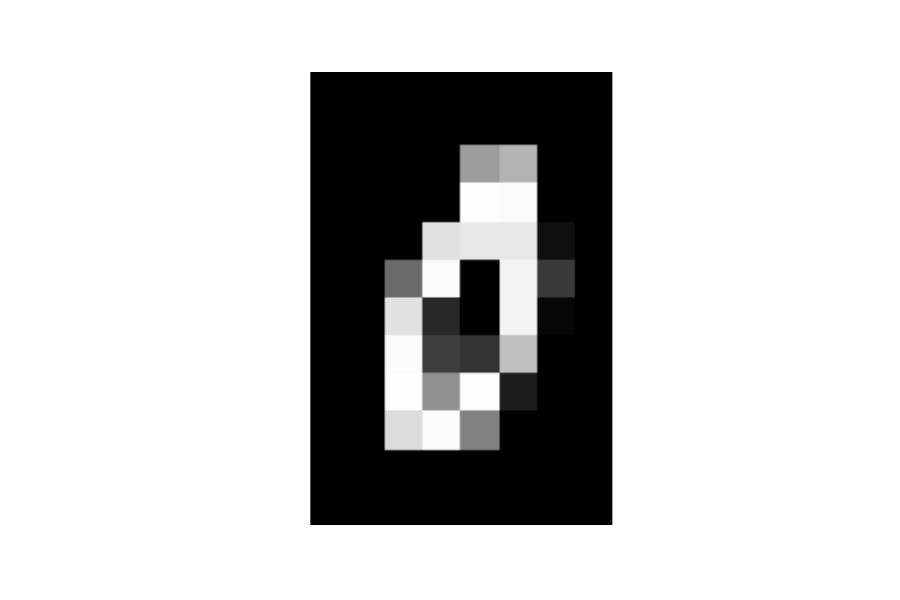}
			\caption{zero}
	\end{subfigure}
	\begin{subfigure}[b]{0.17\linewidth}
		\includegraphics[width=\linewidth]{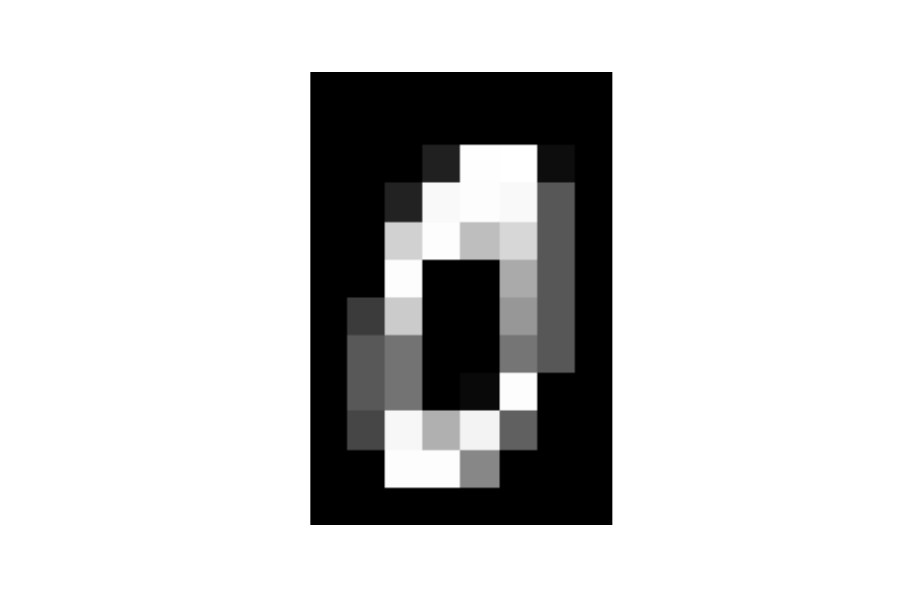}
			\caption{zero}
	\end{subfigure} \\
	\begin{subfigure}[b]{0.17\linewidth}
		\includegraphics[width=\linewidth]{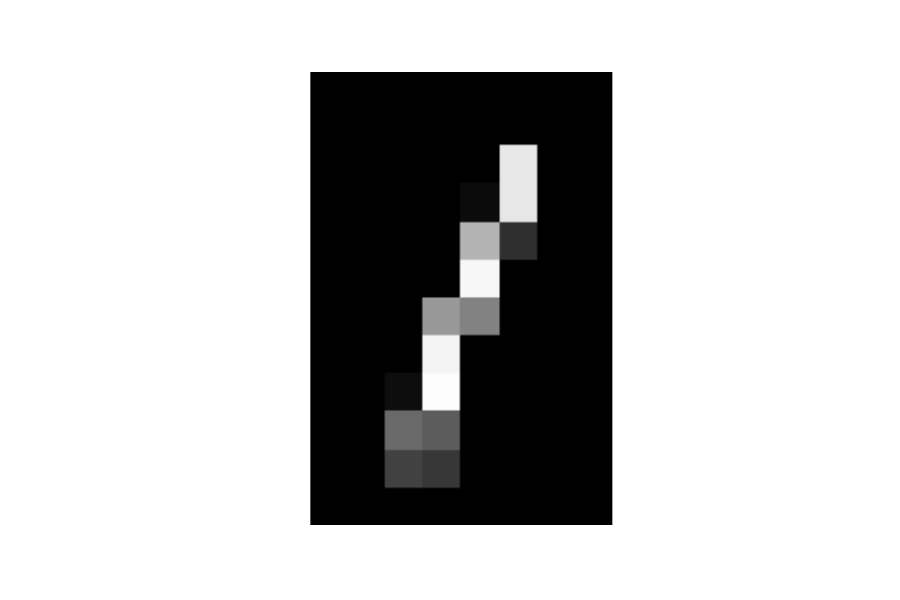}
			\caption{one}
	\end{subfigure}
	\begin{subfigure}[b]{0.17\linewidth}
		\includegraphics[width=\linewidth]{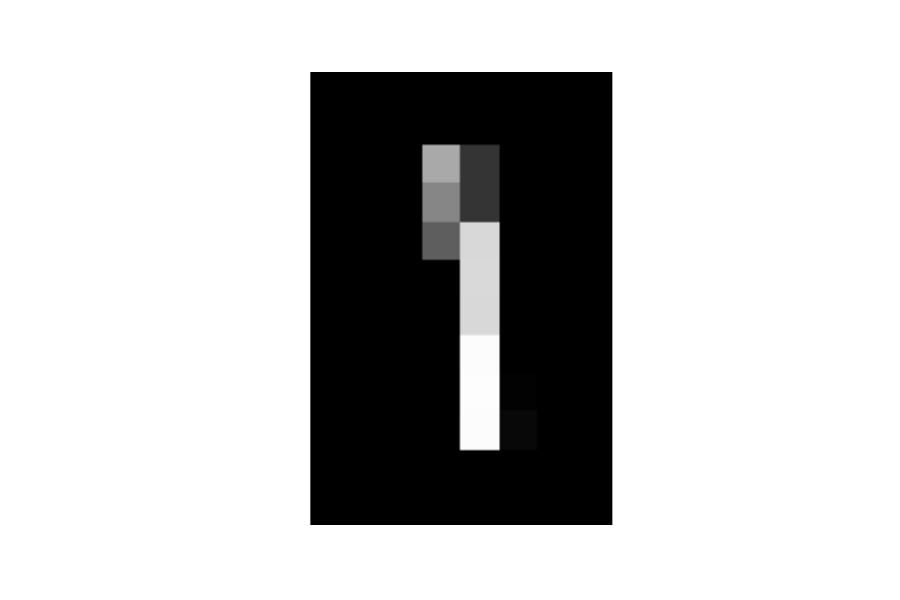}
			\caption{one}
	\end{subfigure}
	\begin{subfigure}[b]{0.17\linewidth}
		\includegraphics[width=\linewidth]{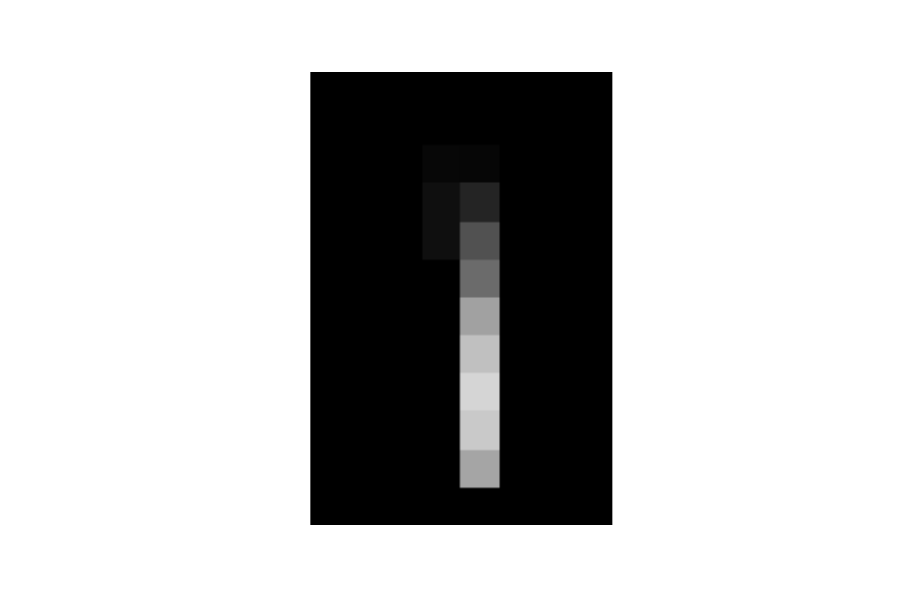}
			\caption{one}
	\end{subfigure}
	\begin{subfigure}[b]{0.17\linewidth}
		\includegraphics[width=\linewidth]{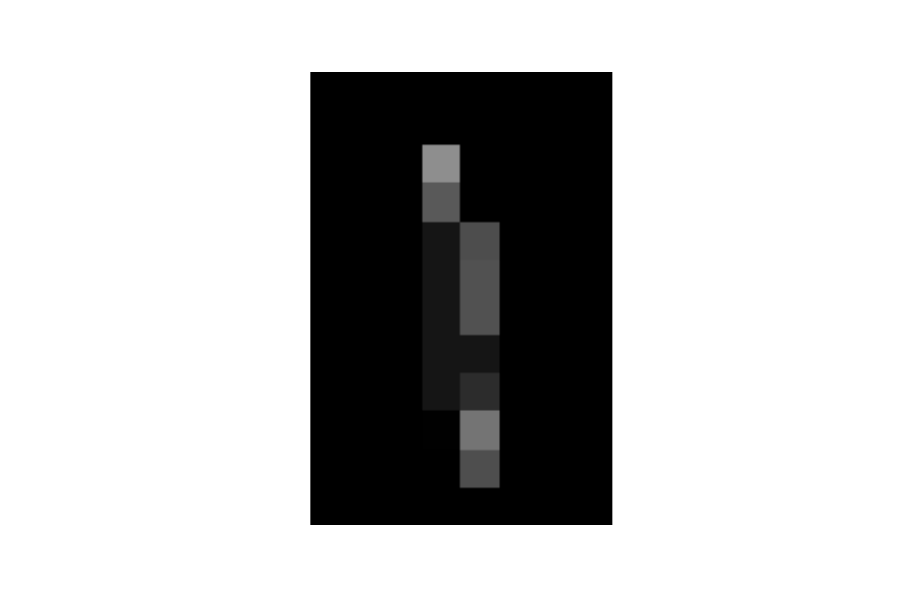}
			\caption{one}
	\end{subfigure}
	\begin{subfigure}[b]{0.17\linewidth}
		\includegraphics[width=\linewidth]{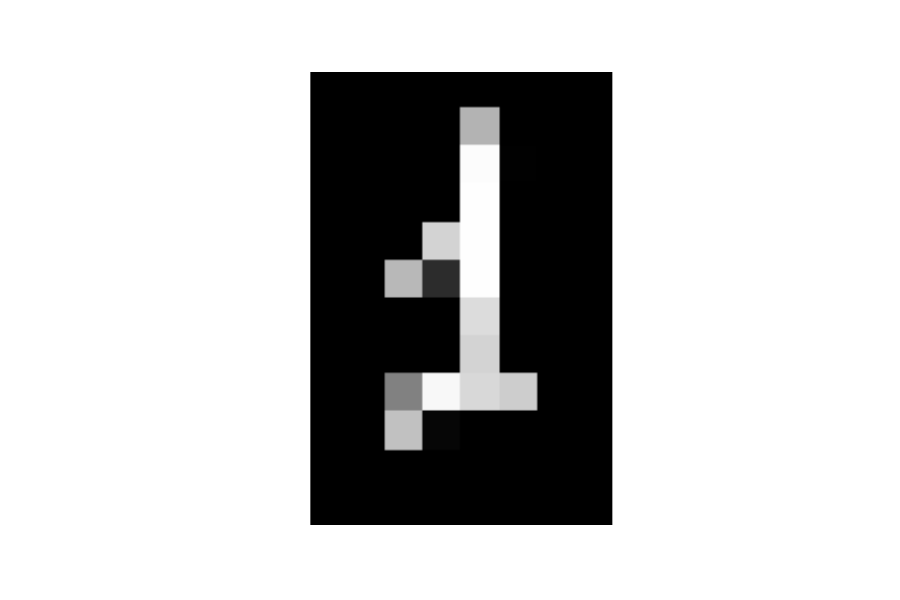}
			\caption{one}
	\end{subfigure}
	\caption{Some images in training set.}
	\label{fig: trainset}
	\end{figure}

We made some numerical simulations to verify the performance of our proposed hybrid quantum-classical normalizing flow model. The model is trained with zeros and ones in MNIST hand-written digits dataset. Some trainset images are shown in Figure \ref{fig: trainset}.

\begin{figure}[h!]
	\centering
	\includegraphics[width=\linewidth]{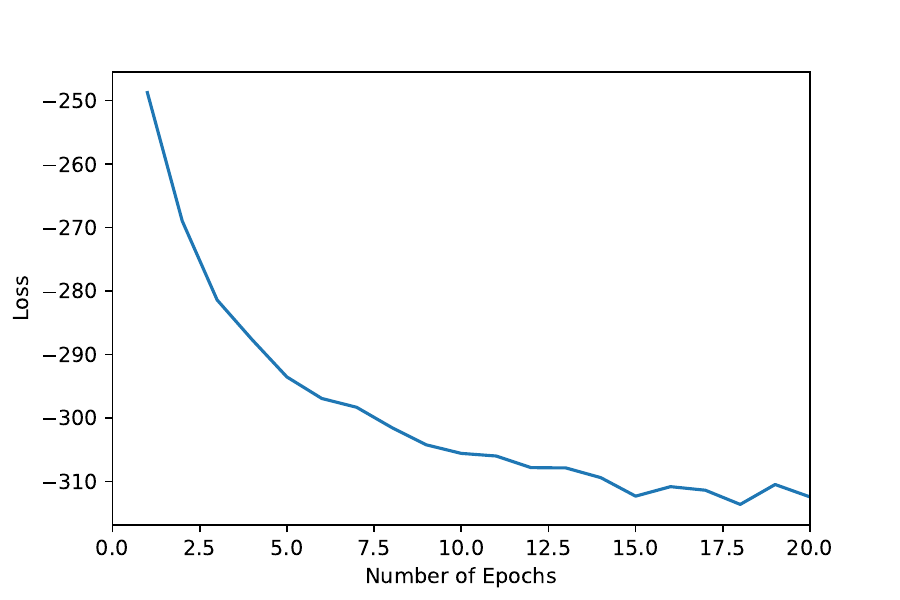}
	\caption{The loss function with training epochs increase.}
	\label{fig:loss}
	\end{figure}

AdamW optimizer \cite{loshchilov2017decoupled} is used to minimize the loss function of the model, and gradient is computed by back-propagation \cite{rumelhart1986learning} (for classical part) and parameter shift rule \cite{schuld2019evaluating} (for quantum part). The loss function goes down with number of epochs goes up, which is shown in Figure \ref{fig:loss}.

\begin{figure}[h!]
	\centering
	\begin{subfigure}[b]{0.17\linewidth}
		\includegraphics[width=\linewidth]{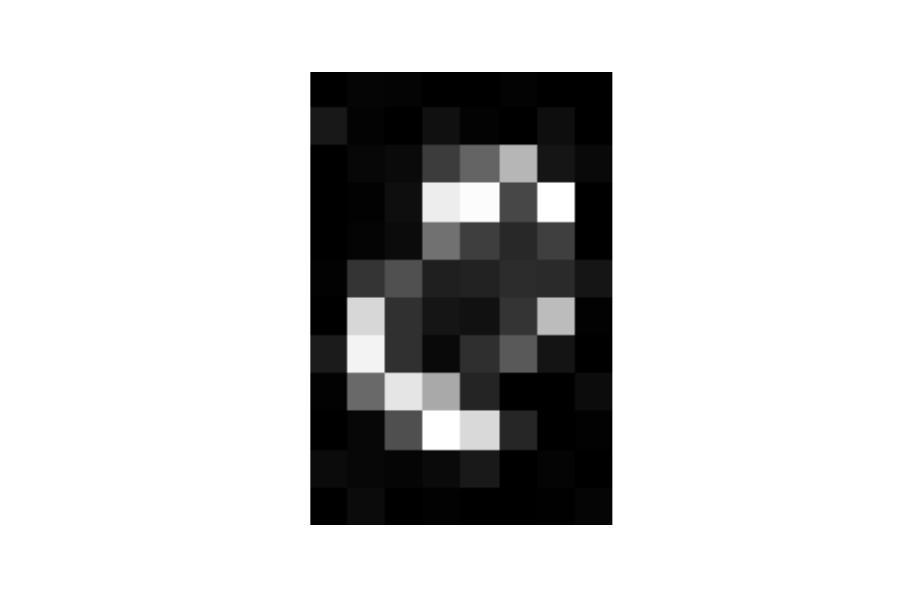}
			\caption{zero}
	\end{subfigure}
	\begin{subfigure}[b]{0.17\linewidth}
		\includegraphics[width=\linewidth]{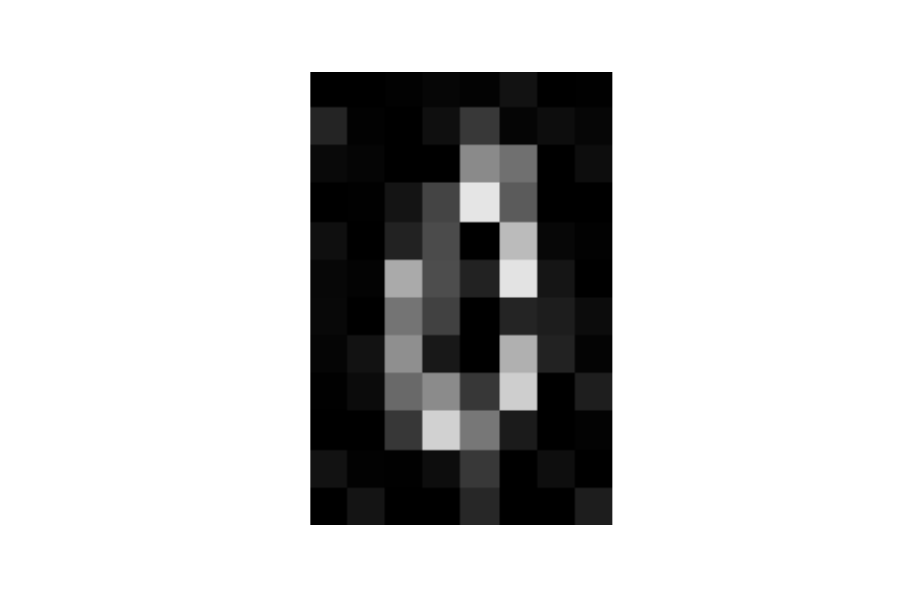}
			\caption{zero}
	\end{subfigure}
	\begin{subfigure}[b]{0.17\linewidth}
		\includegraphics[width=\linewidth]{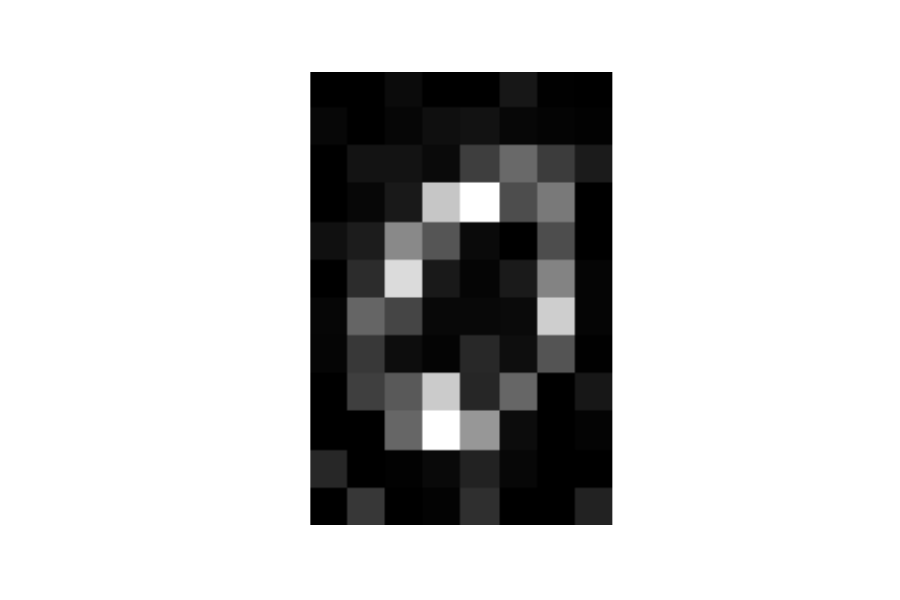}
			\caption{zero}
	\end{subfigure}
	\begin{subfigure}[b]{0.17\linewidth}
		\includegraphics[width=\linewidth]{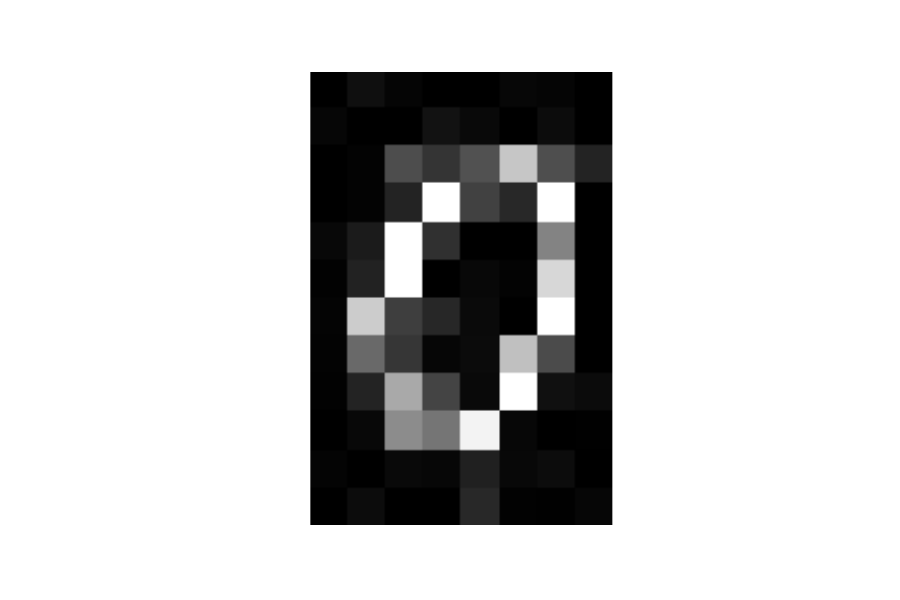}
			\caption{zero}
	\end{subfigure}
	\begin{subfigure}[b]{0.17\linewidth}
			\includegraphics[width=\linewidth]{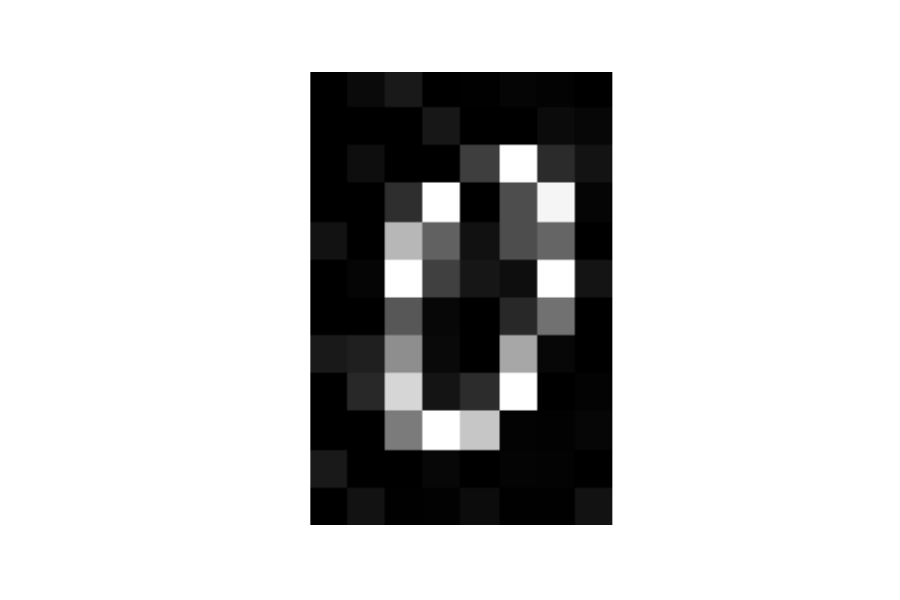}
			\caption{zero}
	\end{subfigure}
	\begin{subfigure}[b]{0.17\linewidth}
		\includegraphics[width=\linewidth]{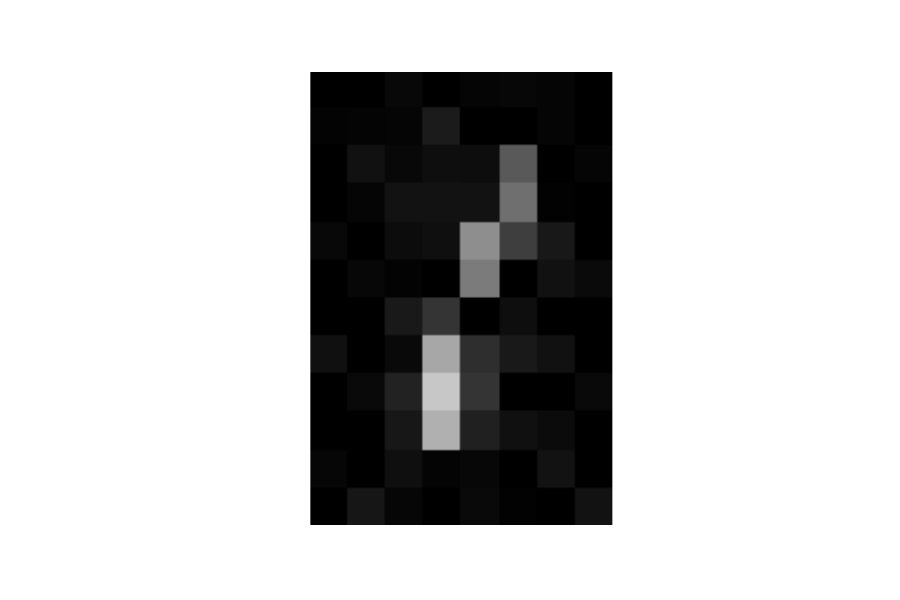}
			\caption{one}
	\end{subfigure}
	\begin{subfigure}[b]{0.17\linewidth}
		\includegraphics[width=\linewidth]{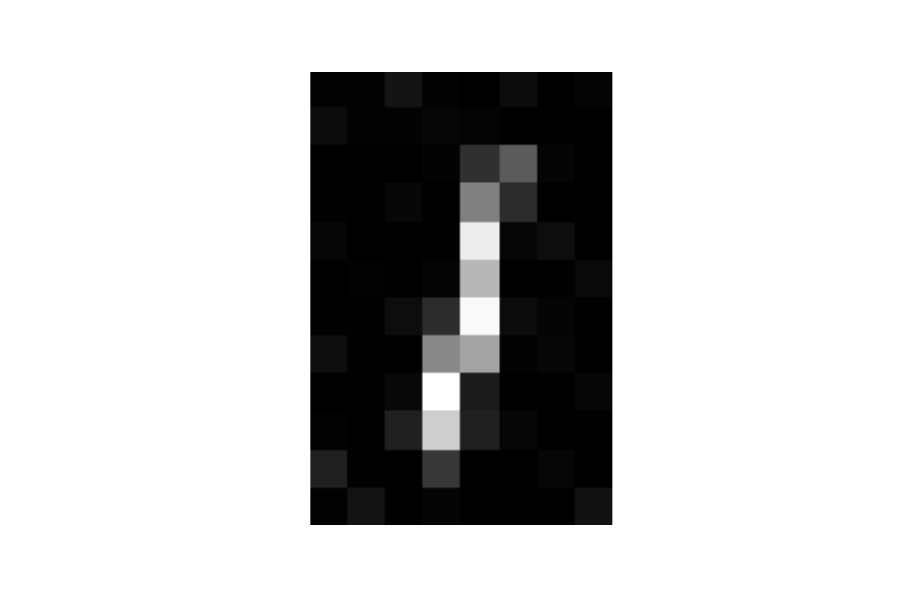}
			\caption{one}
	\end{subfigure}
	\begin{subfigure}[b]{0.17\linewidth}
		\includegraphics[width=\linewidth]{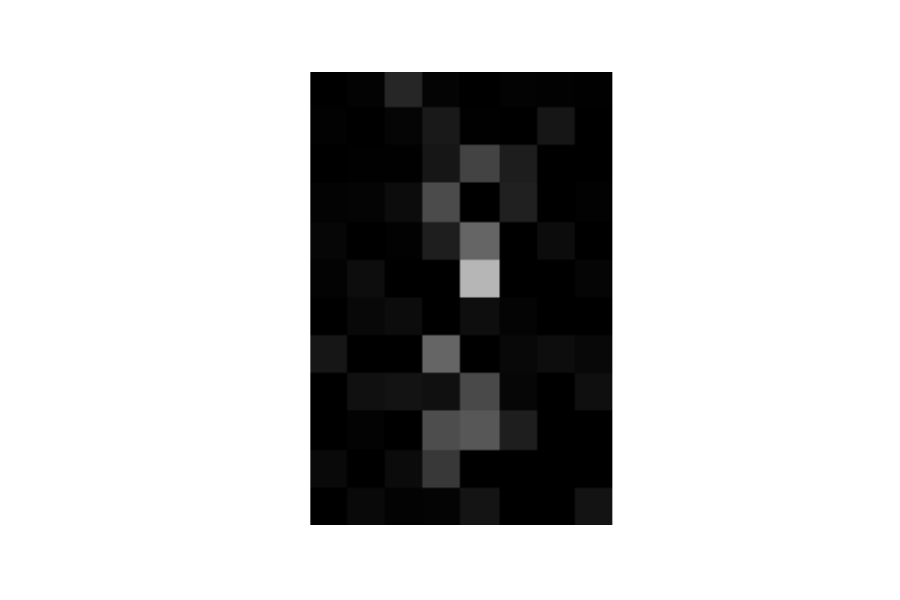}
			\caption{one}
	\end{subfigure}
	\begin{subfigure}[b]{0.17\linewidth}
		\includegraphics[width=\linewidth]{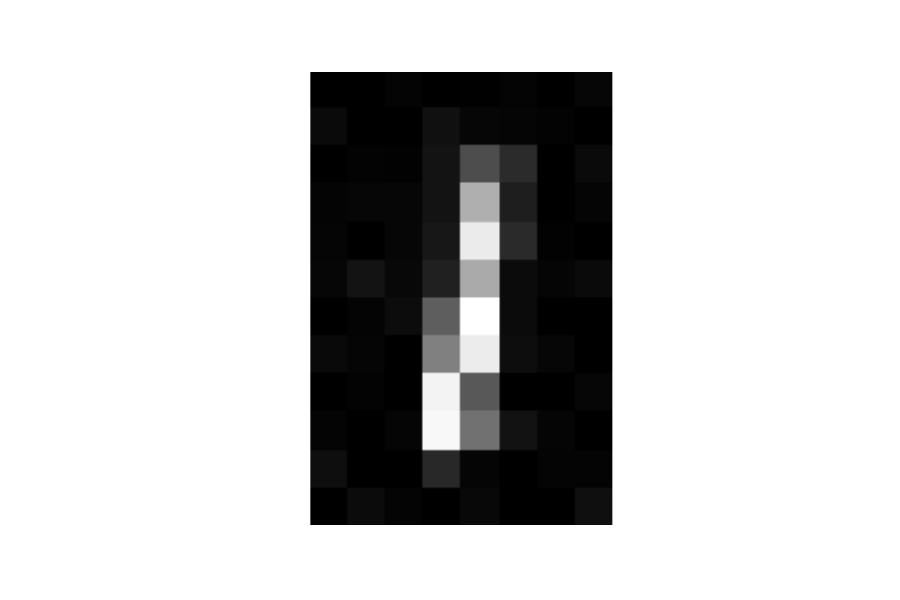}
			\caption{one}
	\end{subfigure}
	\begin{subfigure}[b]{0.17\linewidth}
		\includegraphics[width=\linewidth]{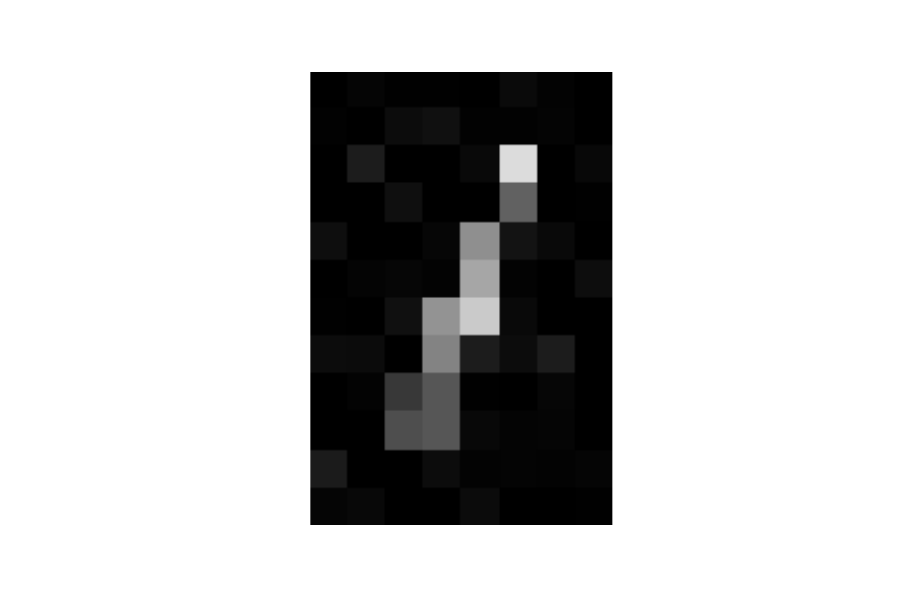}
			\caption{one}
	\end{subfigure}
	\caption{Hand-written digits (zeros and ones) generated by hybrid quantum-classical normalizing flow.}
	\label{fig: generated images}
\end{figure}

After training, our model is able to generate hand-written digits. Though some of the generated images is blurry, some are still able to recognize. Some selected generated images are shown in Figure \ref{fig: generated images}.

\begin{figure}[h!]
	\centering
	\includegraphics[width=\linewidth]{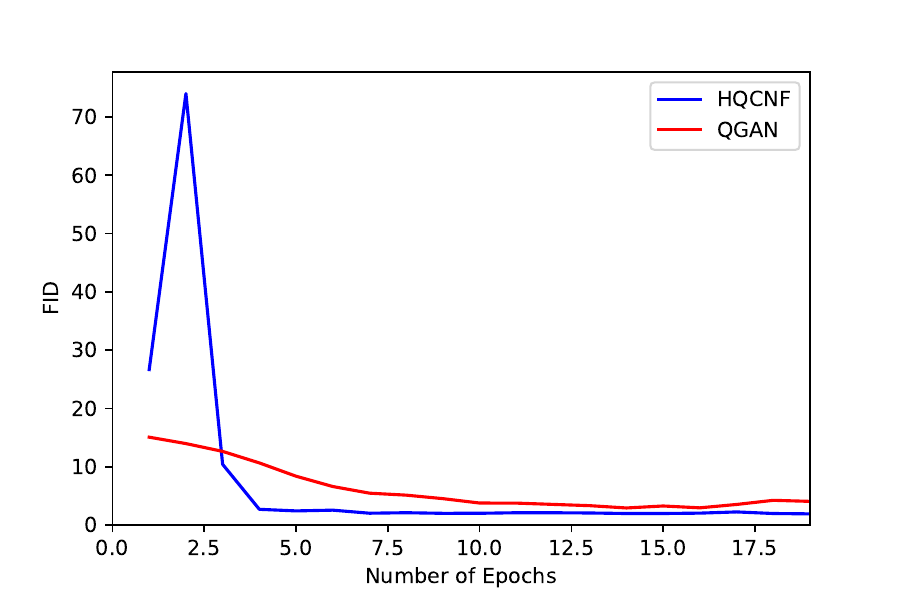}
	\caption{FID score of HQCNF and QGAN during the training process.}
	\label{fig:FID}
\end{figure}

Moreover, we also compute the Frechet Inception Distance (FID) between the generated images and trainset, and compare the result of our model with hybrid quantum GAN model. FID score estimate the distribution gap between the generated images and the trainset; a lower FID score means the distribution of the generated images are close to the trainset, which indicates the corresponding model is better from this certain point of view. The FID score of hybrid quantum-classical normalizing flow is $1.77$, and that of hybrid quantum GAN is $4.80$, which is an evidence to show that our model is better than quantum GAN under this circumstance. Figure \ref{fig:FID} shows the FID score of HQCNF and QGAN during the training process.

The hybrid quantum-classical normalizing flow model proposed in this paper not only demonstrates advantages compared with the currently leading quantum generative models like QGAN, but also has certain advantages over classical generative models. Although quantum generative models can not yet compete with generative models based on classical neural networks due to current development of quantum hardware, our proposed hybrid quantum-classical normalizing flow model still shows several potential advantages. (1) In terms of the number of computing units, the generation process of an $N$-dimensional vector by quantum circuit only requires $n = \log_2 N$ qubits using the method proposed in this paper. This is far less the number of neurons required by classical neural networks to generate vectors of the same dimension. The exponential relationship between the dimension of generated vectors and the number of qubits required enables us to generate high dimensional real-world data with only a few qubits. (2) In terms of the number of model parameters, the quantum circuit part of the model implemented in the paper only uses $40$ parameters for parameterization, which has a significant difference from classical neural networks that usually possess a large number of parameters. Fewer parameters make computation and storage easier. Therefore, using a hybrid quantum-classical normalizing flow model has potential advantages over using classical neural networks alone.

\section{\label{Conclusion}Conclusion}
To sum up, we propose a hybrid quantum-classical normalizing flow model which can tackle real world problems such as generating handwritten digits with NISQ devices. The model extends the normalizing flow framework to quantum configurations while keeping the main features of classical normalizing flows such as nonlinearity and invertibility. We calculate the FID scores of our model and QGAN, proving our model has a better performence. We also compare our model with classical generative models, and shows the potential advantages of our hybrid quantum classical normalizing flow model.

\end{document}